\documentclass[10pt,a4paper]{article}
\usepackage[utf8]{inputenc}
\usepackage[T1]{fontenc}
\usepackage{amsfonts}
\usepackage{amsmath}
\usepackage{amsthm}
\usepackage{amssymb}
\usepackage[dvips]{graphicx}
\usepackage[normalem]{ulem}
\usepackage{xcolor}
\usepackage{tikz}
\usepackage{float}
\usepackage[all,cmtip]{xy}
\usepackage{arydshln}
\usepackage{authblk}
\usepackage{hyperref}

\newtheorem{theorem}{Theorem}[section]
\newtheorem*{theorem*}{Theorem}

\newtheorem*{lemma*}{Lemma}

\newtheorem*{conjecture*}{Conjecture}

\theoremstyle{definition}
\newtheorem{definition}[theorem]{Definition}

\theoremstyle{remark}
\newtheorem{remark}[theorem]{Remark}

\newcommand{\X}{X+\xi}

\newcommand{\TM}{TE \oplus T^*E}

\newcommand{\g}{\mathfrak{g}}

\newcommand{\la}{\left\langle}
\newcommand{\ra}{\right\rangle}

\newcommand{\tB}{\tilde{B}}
\newcommand{\tgi}{\tilde{g}^{-1}}
\newcommand{\tg}{\tilde{g}}

\numberwithin{equation}{section}

\begin{document}
\title{T-duality of emergent gravities on nilmanifolds}
\author[1]{Raju Roychowdhury}
\author[2]{Leonardo Soriani}
\affil[1]{Department of Physics and Computer Science, Dayalbagh Educational Institute, Agra - 282005, India\\
raju.roychowdhury@gmail.com}

\affil[2]{Department of Mathematics, State University of Maringá, Brazil\\
leo.soriani@gmail.com}
\maketitle

\begin{abstract}

We study the transport of generalized metrics between topological T-dual 
nilmanifolds through a Lie algebraic point of view. Emergent gravities are 
generalized metrics with symplectic B-fields. But this additional property
might not be preserved by the aforementioned transport. We describe a 
necessary condition for it to happen and provide working examples on 
self-T-dual nilmanifolds with zero $H$-flux in both 4 and 6 dimensions. 
We also discuss how this procedure fails in the presence of a non-zero $H$-
flux.
\end{abstract}


\section{Introduction}

\subsection{Emergent Gravity} 
It is well known by now that 
the notion of spacetime cannot be a fundamental 
idea due to the failure of a perfect marriage 
between gravity and quantum mechanics.
For arbitrarily smaller length's 
scale of the order of $10^{-33}$ cm, it is hard to 
construct an operational meaning of distance 
leading to the fact that many fundamental 
principles of physics namely those involving 
spacetime symmetry turns out to be an approximate 
one, and there are valid indications of fundamental 
limitations to quantum mechanics in both the early 
and late universe, thus calling for yet another 
deeper structure involving new physical and 
mathematical ideas to save us from the doomed 
idea of spacetime geometry\cite{Nima}. 
As gravity is intricately connected to the idea of 
spacetime; in conformity with the new mathematical 
framework for spacetime, we must root for a 
picture where gravity emerges out of a more 
fundamental paradigm. Emergent Gravity is 
a key player in this quest. 

If at microscopic scales spacetime is viewed as a fabric 
dressed with a noncommutative (NC) structure then 
it necessarily implies emergent spacetime \cite{Yang:2006dk}.
The emergent gravity from NC U(1) gauge theory is the 
large $\mathcal{N}$ duality and this picture accounts 
for a background-independent formulation of quantum 
gravity \cite{Yang:2008fb, Yang:2016ghr} which we 
call emergent theory of gravity. Further interested readers
are referred to the following review articles and papers 
\cite{Lee:2010zf, Yang:2011bd,LRY} to 
explore more about the formulation of emergent gravity, 

In a recent work, \cite{BR} by one of the authors, a T duality 
framework for emergent gravity was laid for the torus
fibrations.  While limiting us to the case of flat 
spacetime (torus fiber with a point base) we were able to
find a T-dual avatar of emergent gravity. For a generic 
non-flat spacetime with a nontrivial base manifold and a 
nontrivial fiber, we could not find one due to the 
limitations imposed by the appearance of 3-form H-flux. 
The present work is an improvement over it in the sense 
that it deals with a completely general case on a nilmanifold 
with $B$ fields that are symplectic both on the base and the fiber. 
We provide examples of the T-duality of emergent gravity on self-T-dual 
nilmanifolds with zero flux. For non-zero flux, we could only show in a 
specific scenario that such a duality of emergent gravity cannot happen.

\subsection{Summary of Results:}
The main findings of our article are as follows. 
Using the Cavalcanti-Gualtieri point of view 
\cite{CG} we can always transport a generalized 
metric and find a geometric T dual generalized 
metric on the other side. But emergent gravity is not "any" 
generalized metric, albeit of a particular kind with a symplectic $B$ 
field and one is not guaranteed to get a dual copy of the emergent 
gravity through the $\varphi$ map unless one imposes some restriction
on the $B$ field such that the dual $B$ also turns out to be symplectic, 
as well, as in conformity with the demands made by emergent gravity. 
Our results are valid in general for any nilmanifold where you have a 
2 form $B$ field which is symplectic both on the fiber and the base 
and for the sake of computational ease, we have tested our condition 
against a diagonal metric $g$ and a generic anti-symmetric $B$ field. 
This situation only occurs for T-dual nilmanifolds with no flux on 
both sides, which implies that they are trivial torus bundles.

The plan of the paper is as follows. In Sec.2 we briefly introduce 
the idea of generalized geometry, topological T duality, and Nil 
manifolds followed by a quick introduction to the recipe provided by
Cavalcanti and Gualtieri on how to build the duality map. Sec.3 
contains our general results for the dual data on Nilmanifolds given 
the initial $(g, B)$ pair, including a criterion that $B$ field has to 
satisfy for its dual to have any chance to be symplectic. In Sec.4 we 
provide explicit examples of T dual emergent gravities in torus and 
non-tori cases worked out in dimensions 4 and 6. Sec. 5 discusses the issue
of non-zero flux and how it curbs the possibility of realizing a 
topological dual of emergent gravity using the Cavalcanti-Gualtieri 
prescription. Finally, in Sec.6 we summarize our results and conclude with
future directions.

\section{Preliminaries}

\subsection{Generalized geometry} This is a quick recap of the facts about generalized complex structures that we will need. Check \cite{Gua} for details. 

Let $E$ be a smooth $n$-dimensional manifold and $H\in \varOmega^3(E)$  be a closed $3$-form. The sum of the tangent and cotangent bundle comes with a natural symmetric bilinear form with signature $(n,n)$ and bracket (the Courant bracket) given by $$\langle X+\xi,Y+\eta \rangle=\frac{1}{2}(\eta(X)+\xi(Y)),$$ $$[\X,Y+\eta]_H=[X,Y]+\mathcal{L}_X\eta-i_Yd\xi+i_Xi_YH.$$

\begin{definition}
A generalized metric on $(E,H)$ is an orthogonal automorphism $\mathcal{G}:\TM \to \TM$ such that  
\begin{itemize}
\item $\mathcal{G}$ is orthogonal;
\item $\mathcal{G}$ is self-adjoint;
\item $\la \mathcal{G}(X+\xi), X+\xi\ra \geq 0$
\end{itemize} 
\end{definition}

The easiest examples of such structures come from Riemannian metrics $g$ on $E$: then $\mathcal{G}_g$ defined below (viewing $g$ as a map $TE \to T^*E$) is a generalized metric on $E$.

$$\mathcal{G}_g:=\left(\begin{array}{cc}
0 & g^{-1}\\
g & 0
\end{array}\right)$$

It follows from the definition that $\mathcal{G}^2=1$, therefore we can consider $C_{\pm}$ the $\pm 1$-eigenspaces of $\mathcal{G}$. $C_+ \subset TE \oplus T^*E$ can be seen as the graph of a linear map $TE \to T^*E$ which we can decompose in its symmetric part $g \in Sym^2 T^*E$ and skew-symmetric part $B \in \Lambda^2 T^*E$. Similar observations for $C_-$ leads to the following description of $C_\pm$:
$$C_\pm=\{X+B(X,\cdot)\pm g(X,\cdot), X \in TE\} $$Therefore a generalized metric $\mathcal{G}$ is equivalent to a pair $(g,B)$ where $g$ is a Riemannian metric and $B$ is a $2$-form. One can write $\mathcal{G}$ in matrix form using $g$ and $B$ as follows (again, viewing $g$ and $B$ as maps $TE \to T^*E$):
$$\mathcal{G}= \left(\begin{array}{cc}
-g^{-1}B & g^{-1}\\
g-Bg^{-1}B & Bg^{-1}
\end{array}\right)$$

\begin{definition}
A generalized metric $\mathcal{G}=(g,B)$ is called an emergent gravity if $B$ is symplectic.
\end{definition}




\medskip
For a Lie group $G$ one can consider invariant \textbf{generalized metrics}. In this case, we assume $H$ to be a left-invariant closed 3-form on $G$, which is identified with an alternating 3-form on the Lie algebra $\mathfrak{g}$ of $G$. The Lie group also acts by left translations on $TG\oplus T^*G$
$$g\cdot (X+\xi)=(L_g)_*X+ (L_{g^{-1}})^*\xi.$$
A generalized complex structure on $(G,H)$ is said to be left invariant if it is equivariant with respect to this action.

\subsection{$T$-duality}
We will follow the notion of topological $T$-duality for principal torus bundles as defined by Bouwknegt, Evslin, Hannabuss and Mathai \cite{BEM}, \cite{BHM}.

\begin{definition}\label{dual}
Let $E$ and $\tilde{E}$ be principal fiber bundles with a $k$-dimensional torus $T$ as the fiber and over the same base $M$ and let $H \in \varOmega^3(E)$, $\tilde{H} \in \varOmega^3(\tilde{E})$ be closed invariant $3$-forms. Let $E \times_M \tilde{E}$ be the product bundle and consider the diagram
$$\xymatrix{
    &(E \times_M \tilde{E},p^*H-\tilde{p}^*\tilde{H}) \ar[ld]^{p}  \ar[rd]_{\tilde{p}}& \\
(E,H)\ar[dr]&  & (\tilde{E},\tilde{H}).\ar[ld]\\
               & M  & }$$

We define $(E,H)$ and $ (\tilde{E},\tilde{H})$ to be $T$-dual if $p^*H-\tilde{p}^*\tilde{H}=d\rho$, where $\rho \in \varOmega^2(E \times_M \tilde{E})$ is a $2$-form $T\times T$-invariant and non-degenerate on the fibers.
\end{definition}

\ 

We observe that, given a torus bundle $E$, it is not always possible to find a T-dual torus bundle for every $3$-form $H$. A necessary condition is that $H(X,Y,\cdot)=0$ if $X,Y$ are tangent to the fiber.

\ 

Given a pair of $T$-dual torus bundles $(E,H)$ and $ (\tilde{E},\tilde{H})$, Cavalcanti and Gualtieri defined an isomorphism between the space of invariant sections preserving the bilinear form and the Courant bracket:
\begin{equation}\label{phi}
\varphi: (\TM)/T \to (T\tilde{E}\oplus T^*\tilde{E})/T.
\end{equation}

Using $\varphi$ one can transport different kinds of geometric structures between $T$-dual spaces, including generalized metrics:

\begin{theorem}\cite{CG}\label{theocg}
Let $(E,H)$ and $(\tilde{E},\tilde{H})$ be $T$-dual torus bundles. If $\mathcal{G}$ is an invariant generalized metric on $E$ then $$\tilde{\mathcal{G}}:=\varphi\circ \mathcal{G} \circ \varphi^{-1}$$ is an invariant generalized metric on $\tilde{E}$.
\end{theorem}

\subsection{ Nilmanifolds}

A nilmanifold is a compact homogeneous manifold $E=\Lambda\backslash G$ where $G$ is a simply connected nilpotent Lie group $G$ and $\Lambda$ is a discrete cocompact subgroup. 

Nilmanifolds can be seen as principal torus bundles over another nilmanifold: let $A$ be a non-trivial $m$-dimensional central normal subgroup of $G$. Then $T=(\Lambda\cap A)\backslash A$ is an $m$-dimensional torus. Since $A\subset Z(G)$, the center of $G$, one has a right action of $T$ on $E$ 
$$ x\cdot a =\Lambda g\cdot  (\Lambda\cap A) z=\Lambda gz\in E,\qquad \mbox{for  }x=\Lambda g\in E,\; a=(\Lambda\cap A )z\in T.$$ The quotient space $M=E/T$ is diffeomorphic to $\Gamma\backslash N$ where $\Gamma=\Lambda A/A\simeq A/\Lambda \cap A$ is a discrete cocompact group of $N$, thus $M=\Gamma\backslash N$ is a nilmanifold. Therefore $E$ is the total space of the principal bundle $q:E\rightarrow M$ with fiber $T$. We refer to a nilmanifold with this specific torus bundle structure in mind as $(G, A,\Lambda)$.

\medskip

This means that nilmanifolds are good spaces if one is looking for examples of T-duality. One can study the T-duality of nilmanifolds at the Lie algebra level, under some conditions. This Lie algebraic point of view was developed in \cite{BGS} and now we will briefly review their main results. The following is a definition of T-duality for Lie algebras, very similar to (\ref{dual}).

\begin{definition}\label{dualLie}
Consider Lie algebras $\mathfrak{g},\tilde{\mathfrak{g}}$ and abelian ideals $\mathfrak{a}\subset\mathfrak{g},\tilde{\mathfrak{a}}\subset\tilde{\mathfrak{g}}$ such that $\mathfrak{n}:=\mathfrak{g}/\mathfrak{a}=\tilde{\mathfrak{g}}/\tilde{\mathfrak{a}}$, closed $3$-forms $H\in \Lambda^3\mathfrak{g}^*$, $\tilde{H} \in \Lambda^3\tilde{\mathfrak{g}}^*$ and the diagram
$$\xymatrix{
 &(\mathfrak{g}\times_{\mathfrak{n}}\tilde{\mathfrak{g}},p^*H-\tilde{p}^*\tilde{H})\ar[ld]^{p}\ar[rd]_{\tilde{p}} & \\
(\mathfrak{g},H)\ar[dr]&  & (\tilde{\mathfrak{g}},\tilde{H})\ar[ld]\\
               & \mathfrak{n} & }$$
We say that $(\mathfrak{g},H)$ and $(\tilde{\mathfrak{g}},\tilde{H})$ are dual if there exist a 2-form $F$ in $\mathfrak{g}\times_{\mathfrak{n}}\tilde{\mathfrak{g}}$ which is non-degenerate in the fibers such that $p^*H-\tilde{p}^*\tilde{H}=dF$.
\end{definition}

If  two nilmanifolds $E=(G,A,\Lambda)$, $\tilde{E}=(\tilde{G},\tilde{A},\tilde{\Lambda})$ with invariant forms $H\in \Omega^3(E)$, $\tilde{H} \in \Omega^3(\tilde{E})$ are T-dual as in definition (\ref{dual}), then the corresponding Lie algebra data $(\mathfrak{g},H)$ and $(\tilde{\mathfrak{g}},\tilde{H})$ is T-dual as in definition (\ref{dualLie}). Under reasonable assumptions, one can start with a T-duality of Lie algebras and integrate it into the framework of 
T-duality of nilmanifolds.

In the following sections, we will talk about specific nilpotent Lie algebras using the Malcev notation $\mathfrak{g}=(de^1, \dots, de^n)$ where $\{e^1,\dots,e^n\}$ is the dual of a basis of $\mathfrak{g}$. The brackets of $\mathfrak{g}$ can be recovered by the formula $d\alpha(X,Y)=-\alpha([X,Y])$. We usually denote $e^i\wedge e^j$ by $e^{ij}$ Then, for example, $(0,0,e^{12},e^{13})$ represent the $4$-dimensional Lie algebra with dual generated by $e^1, \dots , e^4$ such that $de^1=de^2=0$, $de^3=e^1\wedge e^2$ and $de^4=e^1\wedge e^3$. Bracket-wise this is the Lie algebra generated by $e_1,\dots,e_4$ where $[e_1,e_2]=-e_3$ and $[e_1,e_3]=-e_4$ are the non-zero brackets.

 One of the main results of \cite{BGS}  shows how to build the dual of a given Lie algebra in the sense of \ref{dualLie} (if it is possible to do so). The diagram illustrates this procedure and the fact that the flux on one side of the duality dictates the topology of the torus bundle on the other side and vice-versa.

 \begin{center}
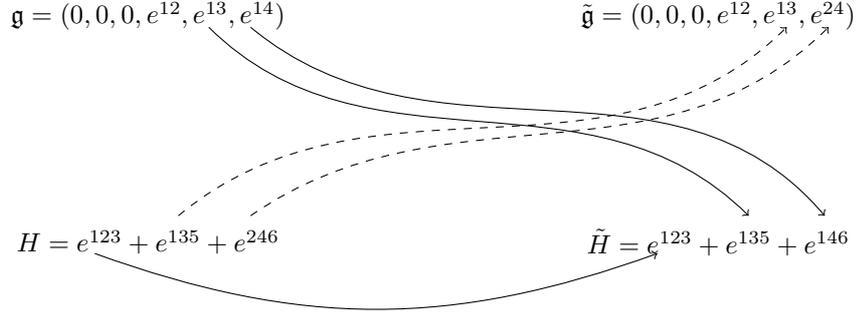
\begin{figure}[H]
\begin{tikzpicture}
\node at (1,0) {$H=e^{123}+e^{135}+e^{246}$};
\node at (8.5,0) {$\tilde{H}=e^{123}+e^{135}+e^{146}$};
\node at (1,3) {$\mathfrak{g}=(0,0,0,e^{12},e^{13},e^{14})$};
\node at (8.5,3) {$\tilde{\mathfrak{g}}=(0,0,0,e^{12},e^{13},e^{24})$};
\draw [->] (1.8,2.85) to [out=-45, in=135](8.9,0.35);
\draw [->] (2.35,2.85) to [out=-40,in=130] (9.9,0.35);
\draw [dashed, ->] (1.4,0.35) to [out=40,in=-135](9.4,2.85);
\draw [dashed, ->] (2.35,0.35) to [out=35,in=-135](9.95,2.85);
\draw [->] (0.3,-0.15) to [out=-20,in=-160](7.7,-0.15);

\end{tikzpicture}
\caption{The diagram shows how to construct the dual of a given admissible triple.}
\end{figure}
\end{center}
 

\begin{remark}\label{zeroHtriv}
The diagram highlights the fact that the flux on one side of the duality dictates the topology of the torus bundle on the other side and vice-versa. In particular, if on one side we have a zero flux the other side has to be a trivial torus bundle.
\end{remark}
The Cavalcanti-Gualtieri map (\ref{phi}) can also be understood on Lie algebras: if $(\mathfrak{g},H)$ and $(\tilde{\mathfrak{g}},\tilde{H})$ are dual then there is an isomorphism $\varphi:\g\oplus \g^*\to\tilde{\g}\oplus \tilde{\g}^*$. In a suitable basis, $\varphi$ has the following form: \begin{equation}\label{phimatrix}\varphi=\left(\begin{array}{cccc}
1_{m\times m} & 0 & 0 & 0\\
0 & 0 & 0 & -1_{k\times k}\\
0 & 0 & 1_{m\times m} & 0\\
0 & -1_{k\times k} & 0 & 0
\end{array}\right). \end{equation}
where $k$ is the dimension of the subgroup $A$, i.e., the dimension of the torus fiber on the corresponding nilmanifold. Then $m$ is the dimension of the base of the torus bundle.

\medskip

\section{Computations on nilmanifolds}



\

\ 
The goal here is to understand when the Cavalcanti-Gualtieri map (\ref{phi}) transforms an emergent gravity into another emergent gravity.

Given a T-dual pair of Lie algebras $(\g,H)$ and $(\tilde{\g},\tilde{H})$, suppose that their torus bundle structure has $k$-dimensional fiber and $m$-dimensional base. For a generalized metric $\mathcal{G}=(g,B)$ on $\g$ and consider its matricial description $$\mathcal{G}= \left(\begin{array}{cc}
-g^{-1}B & g^{-1}\\
g-Bg^{-1}B & Bg^{-1}
\end{array}\right).$$

We can be a bit more detailed in this description taking the fibration into account: $B$ and $g$ are being understood as maps $\g \to \g^*$ which can be split into its basic, fiber and mixed parts

\begin{equation}\label{blocks}B= \left(\begin{array}{cc}
B_b & B_m\\
-B_m^t & B_f
\end{array}\right)\ \ \ g= \left(\begin{array}{cc}
g_b & g_m\\
g_m^t & g_f
\end{array}\right)\end{equation}

Notice that $B_b$ and $g_b$ are $m \times m$ while $B_f$ and $g_f$ are $k\times k$. Also, $B_b$ and $B_f$ are skew-symmetric and $g_b$ and $g_f$ are symmetric.
Here we will take $g_m=0$, so $g$ is diagonal blockwise, which simplifies the computations a little. With these finer blocks, the generalized metric matrix becomes
$$\mathcal{G}=\left(\begin{array}{cccc}
-g_b^{-1}B_b & -g_b^{-1}B_m & g_b^{-1} & 0\\
g_f^{-1}B_m^t & -g_f^{-1} & 0 & g_f^{-1}\\
g_b-B_bg_b^{-1}B_b+B_mg_f^{-1}B_m^t & -B_bg_b^{-1}B_m-B_mg_f^{-1}B_f & B_bg_b^{-1} & B_mg_f^{-1}\\
B_m^tg_b^{-1}B_b+B_fg_f^{-1}B_m^t & g_f+B_m^tg_b^{-1}B_m-B_fg_f^{-1}B_f & -B_m^tg_b^{-1} & B_fg_f^{-1}
\end{array}\right) $$ Observe that the block sizes of this matrix match the block sizes of the matricial description of $\varphi$ (\ref{phimatrix}). Then we find the dual generalized metric $\tilde{\mathcal{G}}$ on $\tilde{\g}$ by matrix multiplication following Theorem \ref{theocg}: $\tilde{\mathcal{G}}=\varphi \cdot \mathcal{G} \cdot \varphi^{-1}$

\begin{equation}\label{tilG}\tilde{\mathcal{G}}=\left(\begin{array}{cccc}
-g_b^{-1}B_b & 0 & g_b^{-1} & g_b^{-1}B_m\\
-B_m^tg_b^{-1}B_b-B_fg_f^{-1}B_m^t & B_fg_f^{-1} & B_m^tg_b^{-1} & g_f+B_m^tg_b^{-1}B_m-B_fg_f^{-1}B_f\\
g_b-B_bg_b^{-1}B_b+B_mg_f^{-1}B_m^t & -B_mg_f^{-1} & B_bg_b^{-1} & B_bg_b^{-1}B_m+B_mg_f^{-1}B_f\\
-g_f^{-1}B_m^t & g_f^{-1} & 0 & -g_f^{-1}B_f
\end{array}\right) \end{equation}

Since $\tilde{\mathcal{G}}$ is a generalized metric, there exist a $2$-form $\tilde{B}$ and a metric $\tilde{g}$ in $\tilde{\g}$ such that

\begin{equation}\label{tilG2}\tilde{\mathcal{G}}= \left(\begin{array}{cc}
-\tilde{g}^{-1}\tilde{B} & \tilde{g}^{-1}\\
\tilde{g}-\tilde{B}\tilde{g}^{-1}\tilde{B} & \tilde{B}\tilde{g}^{-1}
\end{array}\right).\end{equation}

\ 

To extract the dual data $(\tilde{g},\tilde{B})$ on the right side of the duality, we need to compute the inverse of $$\tilde{g}^{-1}=\left(\begin{array}{cc}
g_b^{-1} & g_b^{-1}B_m\\
B_m^tg_b^{-1} & g_f+B_m^tg_b^{-1}B_m-B_fg_f^{-1}B_f
\end{array}\right) $$ which we could not do in the general case, but we will see it in specific examples in the next section. For $$\tilde{B}=\left(\begin{array}{cc}
\tilde{B}_b & \tilde{B}_m\\
-\tilde{B}_m^t & \tilde{B}_f
\end{array}\right)$$ we can do the following: from (\ref{tilG}) and (\ref{tilG2}) we know that

$$ \tilde{B} \cdot \tilde{g}^{-1} = \tilde{B}\tilde{g}^{-1} \iff$$

$$  \left(\begin{array}{cc}
\tilde{B}_b & \tilde{B}_m\\
-\tilde{B}_m^t & \tilde{B}_f
\end{array}\right) \left(\begin{array}{cc}
g_b^{-1} & g_b^{-1}B_m\\
B_m^tg_b^{-1} & g_f+B_m^tg_b^{-1}B_m-B_fg_f^{-1}B_f
\end{array}\right) =\left(\begin{array}{cc}
B_bg_b^{-1} & B_bg_b^{-1}B_m+B_mg_f^{-1}B_f\\
0 & -g_f^{-1}B_f
\end{array}\right)$$ Solving this equation we get

\begin{equation}\label{tilB}\begin{array}{l} \tilde{B}_f=-g_f^{-1}B_f(g_f-B_f g_f^{-1}B_f)^{-1}\\
\tilde{B}_m=-B_m\tilde{B}_f\\
\tilde{B}_b=B_b-\tilde{B}_m B_m^{t}. \end{array}\end{equation}

\ 

Notice that the basic part of $g$ does not affect $\tilde{B}$.

Now suppose that the starting generalized metric $\mathcal{G}=(g, B)$ is an emergent gravity, and let us check if the dual generalized metric $\tilde{\mathcal{G}}$ also represents an emergent gravity, that is if $\tilde{B}$ is a symplectic form.

Let's start by computing $\det \tilde{B}$. Even though we have this rather explicit description of $\tilde{B}$ in (\ref{tilB}), it is a bit cumbersome to try to compute its determinant from this point of view. A smarter way to go about it is to start from 

$$\tilde{B}\tilde{g}^{-1}=\left(\begin{array}{cc}
B_bg_b^{-1} & B_bg_b^{-1}B_m+B_mg_f^{-1}B_f\\
0 & -g_f^{-1}B_f
\end{array}\right) $$ and, since $\tilde{g}$ is a metric $$ \det\tilde{B}\neq 0 \iff \det (\tB\tgi)\neq 0.$$ Computing the latter determinant we have

\begin{eqnarray*} \det(\tB\tgi) & = & \det(B_bg_b^{-1})\cdot\det(-g_f^{-1}B_f)\\
& = & \det(B_b)\cdot\det(g_b^{-1})\cdot\det(-g_f^{-1})\cdot \det(B_f).
\end{eqnarray*}

The blocks $g_b$ and $g_f$ are invertible therefore 

$$\det(\tB\tgi) \neq 0 \iff \det (B_b) \cdot \det (B_f)\neq 0.$$ 

\ 

This means that the $2$-form $\tB$ of the dual generalized metric is non-singular if and only if the basic and fiber parts $B_b$ and $B_f$ of the $2$-form $B$ of the starting generalized metric are non-singular. And since $B_b$ and $B_f$ are skew-symmetric, this can only happen if these blocks are even-dimensional, because if $A$ is an $n\times n$ skew-symmetric matrix then $\det(A)=\det(A^t)=\det(-A)=(-1)^n\det(A)$. We summarize this discussion in the following theorem:

\begin{theorem}\label{theo}
Let $(\g,H)$ and $(\tilde{\g},\tilde{H})$ be T-dual Lie-algebras with T-dual generalized metrics $\mathcal{G}=(g,B)$ and $\tilde{\mathcal{G}}=(\tg,\tB)$. Then $\tB$ is non-singular if and only if $B_b$ and $B_f$ are non-singular (where these are the diagonal blocks of $B$ as in \ref{blocks}). This implies that $B_b$ and $B_f$ are even-dimensional.
\end{theorem}

Now for this $\tilde{B}$ to be symplectic, we still need to check if it is closed. We leave the discussion of closeness in the examples in the next section.


\section{Examples: trivial bundles}

\subsection{Tori}

The torus is the simplest space for us to start. We only care about the even dimensional ones, since the emergent gravities we are looking for require a symplectic form. The $2$-torus does not have enough room for anything interesting to happen in this context, so we will begin with the $4$-torus. The $4$-torus can be seen as a trivial torus bundle in a few different ways: for each $k=0,\dots, 4$ we have $T^4=T^{4-k} \times T^k$  a trivial $T^k$ bundle over $T^{4-k}$. The corresponding Lie algebra data for this torus bundle is the abelian Lie algebra $\mathfrak{g}_4=(0,0,0,0)$ with ideal $\mathfrak{a}_k=\la e_{4-k},\dots,e_4 \ra$. In the absence of flux ($H=0$) these trivial torus bundles are T-dual to themselves (for each $k$).

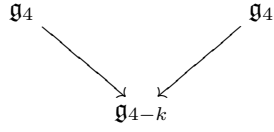
$$\xymatrix{
     \g_4\ar[rd] & &  \tilde{\g}_4\ar[ld]\\
         & \g_{4-k} & }$$

$\g_4=\tilde{\g}_4=(0,0,0,0)=\la e_1,e_2,e_3,e_4 \ra$; $[e_i,e_j]=0$.

$H=\tilde{H}=0$

\ 

From Theorem (\ref{theo}), we know that one cannot expect to have the T-duality of emergent gravity for $k$ odd. For $k=0$, T-duality still makes sense but there is no actual change (the map $\varphi$ is the identity). In the following, we show that for $k=2$ and $k=4$, we can indeed find T-dual emergent gravity.

Since the algebra is abelian, every $2$-form is closed. Therefore we will only need to check if our $2$-forms $B$ are non-degenerate to get symplectic ones. We start with a emergent gravity $\mathcal{G}$ determined a diagonal metric $g$ and any symplectic form $B$:


$$g=\left(\begin{array}{cccc}
g_{11} & 0 & 0 & 0\\
0 & g_{22} & 0 & 0\\
0 & 0 & g_{33} & 0\\
0 & 0 & 0 & g_{44} \end{array}\right),\ \mbox{and}\ B=\left(\begin{array}{cccc}
0 & b_{12} & b_{13} & b_{14}\\
-b_{12} & 0 & b_{23} & b_{24}\\
-b_{13} & -b_{23} & 0 & b_{34} \\
-b_{14} & -b_{24} & -b_{34} & 0 \end{array}\right)$$ with $ (b_{12}b_{34}-b_{13}b_{24}+b_{14}b_{23} )^2\neq 0$. The corresponding generalized metric $\mathcal{G}$ is

$$\mathcal{G}=\left(\begin{array}{cccc;{2pt/2pt}cccc}
0 & -b_{12}g_{11}^{-1} & -b_{13}g_{11}^{-1} & -b_{14}g_{11}^{-1} & g_{11}^{-1}  & 0 & 0 & 0\\
b_{12}g_{22}^{-1} & 0 & -b_{23}g_{22}^{-1} & -b_{24}g_{22}^{-1}  & 0  & g_{22}^{-1} & 0 & 0\\
b_{13}g_{33}^{-1} & b_{23}g_{33}^{-1} & 0 & -b_{34}g_{33}^{-1}  & 0  & 0 & g_{33}^{-1} & 0\\
b_{14}g_{44}^{-1} & b_{24}g_{44}^{-1} & b_{34}g_{44}^{-1} & 0  & 0  & 0 & 0 & g_{44}^{-1}\\ \hdashline[2pt/2pt]
a_{11} & a_{12} & a_{13} & a_{14}  & 0 & b_{12}g_{22}^{-1} & b_{13}g_{33}^{-1} & b_{14}g_{44}^{-1}\\
a_{21} & a_{22} & a_{23} & a_{24}  & -b_{12}g_{11}^{-1} & 0 & b_{23}g_{33}^{-1} & b_{24}g_{44}^{-1}\\
a_{31} & a_{32} & a_{33} & a_{34}  & -b_{13}g_{11}^{-1} & -b_{23}g_{22}^{-1} & 0 & b_{34}g_{44}^{-1}\\
a_{41} & a_{42} & a_{43} & a_{44}  & -b_{14}g_{11}^{-1} & -b_{24}g_{22}^{-1} & -b_{34}g_{33}^{-1} & 0\end{array}\right)$$

$$  \mbox{where}\ \begin{cases}
a_{ii}=g_{ii} + \displaystyle\sum_{j \neq i}b_{ij}^2g_{jj}^{-1}\\
a_{ij}=b_{ik}b_{jk}g_{kk}^{-1}+b_{il}b_{jl}g_{ll}^{-1}, \{i,j\}\cup \{k,l\} =\{1,2,3,4\}\end{cases} $$
(notice that these $a_{ij}$ are symmetric: $a_{ij}=a_{ji}$)

\ 

Recall that the map $\varphi$, for each $k$ is given by 

$$\varphi_k= \left(\begin{array}{cccc}
1_{4-k\times4-k} & 0 & 0 & 0\\
0 & 0 & 0 & -1_{k\times k}    \\
0 & 0 & 1_{4-k\times4-k} & 0\\
0 & -1_{k\times k} & 0 & 0 \end{array}\right)$$

\ 

Now lets finally see how $\mathcal{G}$ transform under T-duality for each of these fibrations: for both $k=2$ and $k=4$ we will compute $\tilde{\mathcal{G}}_k=\varphi_k\circ \mathcal{G}\circ\varphi_k^{-1}$, which is just a shuffling of blocks of $\mathcal{G}$ with some change of signs too.

\

\textbf{k=2}

In this case all the blocks from (\ref{blocks}) are $ 2 \times 2$:

$$g_b=\left(\begin{array}{cc}
g_{11} & 0\\
0 & g_{22}
\end{array}\right), g_f=\left(\begin{array}{cc}
g_{33} & 0\\
0 & g_{44}
\end{array}\right)$$ $$B_b=\left(\begin{array}{cc}
0 & b_{12}\\
-b_{12} & 0
\end{array}\right),B_f=\left(\begin{array}{cc}
0 & b_{34}\\
-b_{34} & 0
\end{array}\right),B_m=\left(\begin{array}{cc}
b_{13} & b_{14}\\
b_{23} & b_{24}
\end{array}\right)$$

\

Following the procedure described in the previous section, the resulting T-dual generalized metric is

$$\tilde{\mathcal{G}}_2=\varphi_2\circ \mathcal{G}\circ\varphi_2^{-1} $$
$$\tilde{\mathcal{G}}_2=\left(\begin{array}{cc|cc;{2pt/2pt}cc|cc}
0 & -b_{12}g_{11}^{-1} & 0 & 0 
& g_{11}^{-1}  & 0 & b_{13}g_{11}^{-1} & b_{14}g_{11}^{-1}\\
b_{12}g_{22}^{-1} & 0 & 0 & 0  
& 0  & g_{22}^{-1} & b_{23}g_{22}^{-1} & b_{24}g_{22}^{-1}\\
\hline
-a_{31} & -a_{32} & 0 & b_{34}g_{44}^{-1}  
& b_{13}g_{11}^{-1} & b_{23}g_{22}^{-1} & a_{33} & a_{34}\\
-a_{41} & -a_{42} & -b_{34}g_{33}^{-1} & 0  
& b_{14}g_{11}^{-1} & b_{24}g_{22}^{-1} & a_{34} & a_{44}\\ 
\hdashline[2pt/2pt]
a_{11} & a_{12} &  -b_{13}g_{33}^{-1} & -b_{14}g_{44}^{-1} 
& 0 & b_{12}g_{22}^{-1} & -a_{13} & -a_{14}\\
a_{21} & a_{22} &  -b_{23}g_{33}^{-1} & -b_{24}g_{44}^{-1} 
& -b_{12}g_{11}^{-1} & 0 & -a_{23} & -a_{24}\\
\hline
-b_{13}g_{33}^{-1} & -b_{23}g_{33}^{-1} & g_{33}^{-1} & 0  
& 0 & 0 & 0 & -b_{34}g_{33}^{-1}\\
-b_{14}g_{44}^{-1} & -b_{24}g_{44}^{-1} & 0 & g_{44}^{-1}  
& 0 & 0 & b_{34}g_{44}^{-1} & 0\end{array}\right)$$

In this matrix, the dotted lines refer to the splitting $\g \oplus \g^*$ while the solid lines represent the base and fiber of the fibration.
\ 

Such generalized metrics can always be seen as coming from a metric $\tilde{g}_2$ and $2$-form $\tilde{B}_2$

$$\tilde{\mathcal{G}}_2=\left(\begin{array}{cc}
-\tilde{g}_2^{-1}\tilde{B}_2 & \tilde{g}_2^{-1}\\
\tilde{g}_2-\tilde{B}_2\tilde{g}_2^{-1}\tilde{B}_2 & \tilde{B}_2\tilde{g}_2^{-1} \end{array}\right),$$

Since $$ \det\tilde{B}_2\tilde{g}_2^{-1}=\dfrac{b_{12}^2b_{34}^2}{g_{11}g_{22}g_{33}g_{44}} $$ and $\det \tilde{g}_2^{-1} \neq 0$ we can already tell that, if $b_{12}^2b_{34}^2 \neq 0$, then $\tilde{B}_2$ is non-degenerate, therefore symplectic. That is, if 
$b_{12} \neq 0$ and $b_{34}\neq 0$ then the dual generalized metric $\tilde{\mathcal{G}}_2=(\tilde{g}_2,\tilde{B}_2)$ is also an emergent gravity. This is a realization of Theorem (\ref{theo}) in this specific example.\\

\vspace{0.5cm}

The following table shows this procedure for some specific choices of $(g, B)$.

\begin{center}
\begin{tabular}{| c | c | } 
  \hline
  
 $ \begin{array}{c} g=e^1e^1+e^2e^2+e^3e^3+e^4e^4\\
  B=e^1\wedge e^2 + e^3 \wedge e^4 \end{array}$ &  $\begin{array}{c} \tilde{g}=e^1e^1+e^2e^2+\dfrac{1}{2}\tilde{e}^3\tilde{e}^3+\dfrac{1}{2}\tilde{e}^4\tilde{e}^4\\
  \tilde{B}=e^1\wedge e^2 -\dfrac{1}{2} \tilde{e}^3 \wedge \tilde{e}^4 \end{array}$\\[4ex]

  \hline

  $ \begin{array}{c} g=e^1e^1+e^2e^2+e^3e^3+e^4e^4\\
  B=e^1\wedge e^2 + e^3 \wedge e^4 +e^1\wedge e^3\end{array}$ &  $\begin{array}{c} \tilde{g}=\dfrac{3}{2}e^1e^1+e^2e^2+\dfrac{1}{2}\tilde{e}^3\tilde{e}^3+\dfrac{1}{2}\tilde{e}^4\tilde{e}^4-\dfrac{1}{2}e^1\tilde{e}^3\\
  \tilde{B}=e^1\wedge e^2 -\dfrac{1}{2} \tilde{e}^3 \wedge \tilde{e}^4 +\dfrac{1}{2}e^1\wedge \tilde{e}^4\end{array}$\\[4ex]
  \hline
  $ \begin{array}{c} g=e^1e^1+e^2e^2+e^3e^3+e^4e^4\\
  B=e^1\wedge e^2 + e^3 \wedge e^4 +e^2\wedge e^4\end{array}$ &  $\begin{array}{c} \tilde{g}=e^1e^1+\dfrac{3}{2}e^2e^2+\dfrac{1}{2}\tilde{e}^3\tilde{e}^3+\dfrac{1}{2}\tilde{e}^4\tilde{e}^4-\dfrac{1}{2}e^2\tilde{e}^4\\
  \tilde{B}=e^1\wedge e^2 -\dfrac{1}{2} \tilde{e}^3 \wedge \tilde{e}^4 -\dfrac{1}{2}e^2\wedge \tilde{e}^3\end{array}$\\[4ex] 
  \hline
  $ \begin{array}{c} g=e^1e^1+e^2e^2+e^3e^3+e^4e^4-\dfrac{1}{2}e^1e^2\\
  B=e^1\wedge e^2 + e^3 \wedge e^4 \end{array}$ &  $\begin{array}{c} \tilde{g}=e^1e^1+e^2e^2+\dfrac{1}{2}\tilde{e}^3\tilde{e}^3+\dfrac{1}{2}\tilde{e}^4\tilde{e}^4-\dfrac{1}{2}e^1e^2\\
  \tilde{B}=e^1\wedge e^2 -\dfrac{1}{2} \tilde{e}^3 \wedge \tilde{e}^4 \end{array}$\\[4ex]
  \hline
  $ \begin{array}{c} g=e^1e^1+e^2e^2+e^3e^3+e^4e^4-\dfrac{1}{2}e^3e^4\\
  B=e^1\wedge e^2 + e^3 \wedge e^4 \end{array}$ &  $\begin{array}{c} \tilde{g}=e^1e^1+e^2e^2+\dfrac{4}{7}\tilde{e}^3\tilde{e}^3+\dfrac{4}{7}\tilde{e}^4\tilde{e}^4+\dfrac{2}{7}\tilde{e}^3\tilde{e}^4\\
  \tilde{B}=e^1\wedge e^2 -\dfrac{4}{7} \tilde{e}^3 \wedge \tilde{e}^4 \end{array}$\\[4ex] 
  \hline
   $ \begin{array}{c} g=e^1e^1+e^2e^2+e^3e^3+e^4e^4-\dfrac{1}{2}e^3e^4\\
  B=e^1\wedge e^2 + e^3 \wedge e^4 +e^1\wedge e^3\end{array}$ &  $\begin{array}{c} \tilde{g}=\dfrac{11}{7}e^1e^1+e^2e^2+\dfrac{4}{7}(\tilde{e}^3\tilde{e}^3+\tilde{e}^4\tilde{e}^4)\\-\dfrac{2}{7}(2e^1\tilde{e}^3+e^1\tilde{e}^4-\tilde{e}^3\tilde{e}^4)\\
  \tilde{B}=e^1\wedge e^2 -\dfrac{4}{7} \tilde{e}^3 \wedge \tilde{e}^4+\dfrac{4}{7}e^1\wedge\tilde{e}^4 \end{array}$\\[6ex] 
  \hline

\end{tabular}
\end{center}

\ \

\textbf{k=4}

\

This is an extreme case with $0$-dimensional base and $4$-dimensional fiber. So the whole space is a fiber and everything changes through T-duality. This scenario has been already studied in \cite{BR}.

Here $\varphi$ takes the simple form $$\varphi_4= \left(\begin{array}{cc}
0 & -1_{4\times4}\\
-1_{4\times4} & 0 \end{array} \right)$$

and the new generalized metric can be expressed as

$$\tilde{\mathcal{G}}_4=\varphi_4\circ\mathcal{G}\circ \varphi_4^{-1} =\left(\begin{array}{cc}
Bg^{-1} & g-Bg^{-1}B\\
g^{-1} & -g^{-1}B \end{array} \right)$$

$$=\left(\begin{array}{cccc;{2pt/2pt}cccc}
0 & b_{12}g_{22}^{-1} & b_{13}g_{33}^{-1} & b_{14}g_{44}^{-1} & a_{11} & a_{12} & a_{13} & a_{14}\\
-b_{12}g_{11}^{-1} & 0 & b_{23}g_{33}^{-1} & b_{24}g_{44}^{-1} & a_{21} & a_{22} & a_{23} & a_{24}\\
b_{13}g_{11}^{-1} & -b_{23}g_{22}^{-1} & 0 & b_{34}g_{44}^{-1} & -a_{31} & a_{32} & a_{33} & a_{34}\\
-b_{14}g_{11}^{-1} & -b_{24}g_{22}^{-1} & -b_{34}g_{33}^{-1} & 0  & a_{41} & a_{42} & a_{43} & a_{44}\\ \hdashline[2pt/2pt]
g_{11}^{-1}  & 0 & 0 & 0 & 0 & -b_{12}g_{11}^{-1} & -b_{13}g_{11}^{-1} & -b_{14}g_{11}^{-1}\\
 0  & g_{22}^{-1} & 0 & 0  & b_{12}g_{22}^{-1} & 0 & -b_{23}g_{22}^{-1} & -b_{24}g_{22}^{-1}\\
0  & 0 & g_{33}^{-1} & 0 &b_{13}g_{33}^{-1} & b_{23}g_{33}^{-1} & 0 & -b_{34}g_{33}^{-1}\\
0  & 0 & 0 & g_{44}^{-1}& b_{14}g_{44}^{-1} & b_{24}g_{44}^{-1} & b_{34}g_{44}^{-1} & 0 \end{array}\right)$$

$$\Rightarrow \tilde{g}_4^{-1}=\left(\begin{array}{cccc}
a_{11} & a_{12} & a_{13} & a_{14}\\
a_{21} & a_{22} & a_{23} & a_{24}\\
a_{31} & a_{32} & a_{33} & a_{34}\\
a_{41} & a_{42} & a_{43} & a_{44}\\ \end{array}\right)$$

Theorem (\ref{theo}) is not very insightful in this situation, because there is no basic part of $B$, everything is on the fiber.


\

\

These results can be generalized for higher even-dimensional tori. See Remark (\ref{rmkT6}) for examples on $T^6$.

\ 

\subsection{non-tori examples}

We can generalize these ideas one step further: let $N$ be any nilmanifold and consider the trivial torus bundle $N\times T^k$. Again, with no flux, these bundles are self-T-dual. From the Lie algebraic point of view, these self-T-dualities are of the form
$$\xymatrix{
     \mathfrak{n}\oplus \mathfrak{a}_k \ar[rd] & &  \mathfrak{n}\oplus \mathfrak{a}_k\ar[ld]\\
         & \mathfrak{n} & }$$ where $\mathfrak{n}$ is a nilpotent Lie algebra and $\mathfrak{a}_k$ is the $k$-dimensional abelian Lie algebra.

If $\mathfrak{n}$ is even dimensional and $k$ is even one can produce other examples of T-duality of emergent gravity. One difference that needs to be noted is that if $\mathfrak{n}$ is not abelian it has left invariant forms that are not closed. One needs to check for the closeness of $B$ to ensure that it is symplectic. 

The only $2$-dimensional nilmanifold is the $2$-torus, therefore to produce new examples one should take $\mathfrak{n}$ to be at least $4$-dimensional. There are $2$ nonabelian $4$-dimensional Lie algebras: $\mathfrak{kt}=(0,0,0,12)$ (denoted this way because the Kodaira-Thurston manifold \cite{Thurston} is a nilmanifold associated to this Lie algebra) and the filiform Lie algebra $\mathfrak{f}_4=(0,0,12,13)$.

Let's see a specific example of this situation with $\mathfrak{n}=\mathfrak{kt}$ and $k=2$. Then we have the $6$-dimensional Lie algebra $\g= \mathfrak{kt} \oplus \mathfrak{a}_2=(0,0,0,12,0,0)$.

$$\xymatrix{
     \g\ar[rd] & &  \tilde{\g}\ar[ld]\\
         & \mathfrak{kt} & }$$

$\g=\tilde{\g}=(0,0,0,12,0,0)=\la e_1,e_2,e_3,e_4,e_5,e_6 \ra$; $[e_1,e_2]=-e_4$

$H=\tilde{H}=0$

\ 

Consider an emergent gravity given by a diagonal metric $g=\sum g_{ii}e_i^2$ and symplectic form $B=e^1\wedge e^4+e^2 \wedge e^3+ e^5\wedge e^6$ (notice that $B$ is closed: any $2$-form $\sum b_{ij}e^i\wedge e^j$ in $\g$ is closed if $b_{34}=0$). Following the same procedure as before we get the dual generalized metric given by $$\tilde{g}=g_{11}e_1^2+g_{22}e_2^2+g_{33}e_3^2+g_{44}e_4^2+\frac{g_{66}}{g_{55}g_{66}+1}e_5^2+\frac{g_{55}}{g_{55}g_{66}+1}e_6^2 $$

$$\tilde{B}=e^1\wedge e^4+e^2 \wedge e^3-\dfrac{1}{g_{55}g_{66}}e^5\wedge e^6 $$ Since the T-dual $2$-form $\tilde{B}$ is closed ($b_{34}=0$) $(\tilde{g},\tilde{B})$ is also an emergent gravity and this represents a working example of T-duality of emergent gravities.


\ 

Now if we take $\mathfrak{n}=\mathfrak{f}_4$ we have the following self T-duality
$$
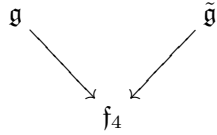
\xymatrix{
     \g\ar[rd] & &  \tilde{\g}\ar[ld]\\
         & \mathfrak{f}_4 & }$$

$\g=\tilde{\g}=(0,0,12,13,0,0)=\la e_1,e_2,e_3,e_4,e_5,e_6 \ra$;
$[e_1,e_2]=-e_3; [e_1,e_3]=-e_4$;
$H=\tilde{H}=0$

\

In this Lie algebra $\mathfrak{g}$ a $2$-form $B=\sum b_{ij}e^i\wedge e^j$ is closed if and only if ${b_{24}=b_{34}=}0$. Therefore choosing $g$ and $B$ as before we produce another example of the T-duality of emergent gravities:
$$ \begin{array}{ccc}
g=\displaystyle\sum_{i=1}^{6}g_{ii}e_i^2 & \iff& \tilde{g}=\displaystyle\sum_{i=1}^{4}g_{ii}e_i^2+\frac{g_{66}}{g_{55}g_{66}+1}e_5^2+\frac{g_{55}}{g_{55}g_{66}+1}e_6^2\\
B=e^1\wedge e^4+e^2 \wedge e^3+ e^5\wedge e^6 & \iff& \tilde{B}=e^1\wedge e^4+e^2 \wedge e^3-\dfrac{1}{g_{55}g_{66}}e^5\wedge e^6
\end{array}$$

We remark that even though the emergent gravities $(g,B)$ and $(\tilde{g},\tilde{B})$ look the same in these two examples, they are not identical. We are using the same notation $e_1,\dots,e_6$ for a basis in different Lie algebras in each example. In general, we can do this sort of calculation starting with a generalized metric $(g,B)$ for any Lie algebra of the form $\mathfrak{n}\oplus \mathfrak{a}$ with $\mathfrak{n}$ nilpotent and $\mathfrak{a}$ abelian. In light of Theorem (\ref{theo}) we can always choose $B$ non-singular with $\tilde{B}$ to be non-singular as well. Then for each Lie algebra $\mathfrak{n}$ we check if both $B$ and $\tilde{B}$ are closed since different Lie algebras can have different closed forms.

\vspace{0.5cm}

While in the previous example the pair of generalized metrics $(g,B)$ and $(\tilde{g},\tilde{B})$ turned out to be emergent gravities on both sides for different choices of $\mathfrak{n}$, this is not always the case. The table below show different choices of $(g,B)$ and the corresponding dual data $(\tilde{g},\tilde{B})$ in $\mathfrak{n}\oplus \mathfrak{a}_2$ for $4$-dimensional $\mathfrak{n}$. We chose each $B$ symplectic in both $\mathfrak{kt} \oplus \mathfrak{a}_2$ and $\mathfrak{f}_4 \oplus \mathfrak{a}_2$ ($b_{24}=b_{34}=0$). The dual data has symplectic $\tilde{B}$ in almost all the rows for both Lie algebras, which characterizes the T-duality of emergent gravities. The only exception is the fifth row, where $\tilde{b}_{24}=\frac{1}{2}$ and $\tilde{b}_{34}=0$: therefore it is T-duality of emergent gravities for $\mathfrak{n}=\mathfrak{kt}$ but not for $\mathfrak{n}=\mathfrak{f}_4$

\begin{center}
\begin{tabular}{| c | c | } 
  \hline
 $ \begin{array}{c} g=e^1e^1+e^2e^2+e^3e^3+e^4e^4+e^5e^5+e^6e^6\\
 \\
  B=e^1\wedge e^4 + e^2 \wedge e^3 + e^5 \wedge e^6 \end{array}$ &  $\begin{array}{c} \tilde{g}=e^1e^1+e^2e^2+e^3e^3+e^4e^4\\ +\dfrac{1}{2}(\tilde{e}^5\tilde{e}^5+\tilde{e}^6\tilde{e}^6)\\
  \tilde{B}=e^1\wedge e^4 + e^2 \wedge e^3 -\dfrac{1}{2} \tilde{e}^5 \wedge \tilde{e}^6 \end{array}$\\[5ex] 
  \hline
  $ \begin{array}{c} g=e^1e^1+e^2e^2+e^3e^3+e^4e^4+e^5e^5+e^6e^6\\
  \\
  B=e^1\wedge e^4 + e^2 \wedge e^3 + e^5 \wedge e^6\\ + e^2 \wedge e^5 \end{array}$ &  $\begin{array}{c} \tilde{g}=e^1e^1+\dfrac{3}{2}e^2e^2+e^3e^3+e^4e^4\\+\dfrac{1}{2}(\tilde{e}^5\tilde{e}^5+\tilde{e}^6\tilde{e}^6)-\dfrac{1}{2}e^2\tilde{e}^5\\
  \tilde{B}=e^1\wedge e^4 + e^2 \wedge e^3  -\dfrac{1}{2} \tilde{e}^5 \wedge \tilde{e}^6 +\dfrac{1}{2}e^2\wedge \tilde{e}^6\end{array}$\\[6ex]
  \hline
  $ \begin{array}{c} g=e^1e^1+e^2e^2+e^3e^3+e^4e^4+e^5e^5+e^6e^6\\ -\dfrac{1}{3} e^5e^6\\
  \\
  B=e^1\wedge e^4 + e^2 \wedge e^3 + e^5 \wedge e^6\end{array}$ &  $\begin{array}{c} \tilde{g}=e^1e^1+e^2e^2+e^3e^3+e^4e^4\\+\dfrac{9}{17}(\tilde{e}^5\tilde{e}^5+\tilde{e}^6\tilde{e}^6)+\dfrac{3}{17}\tilde{e}^5\tilde{e}^6\\
  \\
  \tilde{B}=e^1\wedge e^4 + e^2\wedge e^3 -\dfrac{9}{17}\tilde{e}^5\wedge \tilde{e}^6\end{array}$\\[7ex] 
  \hline
  $ \begin{array}{c} g=e^1e^1+e^2e^2+e^3e^3+e^4e^4+e^5e^5+e^6e^6\\ -\dfrac{1}{3} e^5e^6\\
  \\
  B=e^1\wedge e^4 + e^2 \wedge e^3 + e^5 \wedge e^6\\ + e^2 \wedge e^5  \end{array}$ &  $\begin{array}{c} \tilde{g}=e^1e^1+\dfrac{26}{17}e^2e^2+e^3e^3+e^4e^4\\+\dfrac{9}{17} (\tilde{e}^5\tilde{e}^5+\tilde{e}^6\tilde{e}^6)  -\dfrac{3}{17}(e^2\tilde{e}^6 - \tilde{e}^5\tilde{e}^6 + 3e^2\tilde{e}^5) \\
  \\
   \tilde{B}=e^1\wedge e^4 + e^2\wedge e^3 -\dfrac{9}{17}\tilde{e}^5\wedge \tilde{e}^6\\ +\dfrac{9}{17} e^2\wedge \tilde{e}^6 \end{array}$\\[10ex]
  \hline
  $ \begin{array}{c} g=e^1e^1+e^2e^2+e^3e^3+e^4e^4+e^5e^5+e^6e^6\\
  \\
  B=e^1\wedge e^4 + e^2 \wedge e^3 + e^5 \wedge e^6\\ + e^2 \wedge e^5 + e^4 \wedge e^6 \end{array}$ &  $\begin{array}{c} \tilde{g}=e^1e^1+\dfrac{3}{2}e^2e^2+e^3e^3+\dfrac{3}{2}e^4e^4\\+\dfrac{1}{2} (\tilde{e}^5\tilde{e}^5+\tilde{e}^6\tilde{e}^6) -\dfrac{1}{2}(e^2\tilde{e}^5 + e^4\tilde{e}^6) \\
  \\
  \tilde{B}=e^1\wedge e^4 + e^2\wedge e^3 -\dfrac{1}{2}\tilde{e}^5\wedge \tilde{e}^6\\ +\dfrac{1}{2}(  e^2\wedge e^4 - e^2 \wedge \tilde{e}^6 + e^4 \wedge \tilde{e}^5) \end{array}$\\[10ex] 
  \hline
   $ \begin{array}{c} g=e^1e^1+e^2e^2+e^3e^3+e^4e^4+2e^5e^5+e^6e^6\\+e^5e^6\\
   \\
   B=e^1\wedge e^4 + e^2 \wedge e^3 + e^5 \wedge e^6\\ + e^1 \wedge e^5 + e^1 \wedge e^6 + e^3 \wedge e^6\end{array}$ &  $\begin{array}{c} \tilde{g}=\dfrac{3}{2}e^1e^1+e^2e^2+2e^3e^3+e^4e^4+\dfrac{1}{2} \tilde{e}^5\tilde{e}^5+\tilde{e}^6\tilde{e}^6\\ -\dfrac{1}{2}(\tilde{e}^5\tilde{e}^6 + e^1\tilde{e}^6 -e^1e^3 - e^3\tilde{e}^5 +2e^3\tilde{e}^6) \\
   \\
  \tilde{B}=e^1\wedge e^4 + e^2\wedge e^3 -\dfrac{1}{2}\tilde{e}^5\wedge \tilde{e}^6\\ -\dfrac{1}{2} (e^3\wedge \tilde{e}^5- e^1 \wedge \tilde{e}^6 + e^1 \wedge \tilde{e}^5)\end{array}$\\[10ex] 
  \hline
  
\end{tabular}
\end{center}

\begin{remark}\label{rmkT6}
The table above can also be seen as an example of the T-duality of emergent gravities on $T^6=T^4 \times T^2$ by taking $\mathfrak{n}=\mathfrak{a}_4$.
\end{remark}

\section{Non zero flux}

\ 


All the examples from the previous sections are on self-T-dualities, where the $H$-flux is zero on both sides. We could not find working examples with non-zero flux (in at least one of the sides). Here we discuss this problem and examine one example a little closer.

 Recall from Remark (\ref{zeroHtriv}) that non-zero flux on one side of the duality forces the other side to be a non-trivial torus bundle. Regarding Lie algebras, a non-trivial torus bundle means a Lie algebra with more non-zero brackets. This then gives rise to more non-closed differential forms. In the end, the more complicated the topology of the torus bundle structure of the nilmanifold, the harder it is to find closed (and therefore, symplectic) 2-forms. Let us investigate this situation in the following example.

Consider the T-duality between the $4$-torus and the Kodaira-Thurston nilmanifold, which can be stated using the Lie algebra point of view in the following way:

$$\xymatrix{
  (\g_4,\tilde{H})   \ar[rd] & &  (\mathfrak{kt},H)\ar[ld]\\
         & \g_2 & }$$

         with $\g_4=(0,0,0,0)$, $\mathfrak{kt}= (0,0,0,12)$, $\g_2=(0,0)$

         $H=e^1\wedge e^2 \wedge e^4$, $\tilde{H}=0$.

       \ 
       
Given the diagonal metric $g=e_1^2+e_2^2+e_3^2+e_4^2$ on the left side, let's see if we can find a $B=\sum b_{ij}e^i\wedge e^j$ such that $(g,B)$ is an emergent gravity and the corresponding dual generalized metric $(\tilde{g},\tilde{B})$ is also an emergent gravity. 

\ 

From Theorem (\ref{theo}) we know that if the coefficients $b_{12}$ and $b_{34}$ are non-zero, $\tilde{B}$ is non-singular. Now we need to check if  $\tilde{B}=\sum \tilde{b}_{ij}e^i\wedge e^j$ is closed. But notice that on $\mathfrak{kt}$ we have $de^4=e^1\wedge e^2$ and the only summand of $\tilde{B}$ that is not closed is $e^3\wedge e^4$: $d(e^3\wedge e^4)=e^1\wedge e^2 \wedge e^3$. Therefore, $\tilde{B}$ is closed if and only if $\tilde{b}_{34}=0$.

But if  $b_{34}\neq 0$ on the left side, then we get $\tilde{b}_{34}\neq 0$ on the right side too. Indeed, following (\ref{tilB}) we have $$\tilde{B}_f=\left(\begin{array}{cc}
0 & -\dfrac{b_{34}}{1+b_{34}^2}\\
\dfrac{b_{34}}{1+b_{34}^2} & 0 \end{array}\right), $$ that is, $\tilde{b}_{34}=-\dfrac{b_{34}}{1+b_{34}^2}$. Therefore, there is no duality of emergent gravity in this T-duality picture. Similar problems arise when one attempts to find examples like this in $6$ dimension. 

\section{Conclusion}
The salient points of our results can be summarized as follows:\\
In section $3$ we described the transport of generalized metrics between T-dual nilmanifolds using the Cavalcanti-Gualtieri map \cite{CG} and the Lie algebraic point of view of \cite{BGS}. By breaking down the matrices related to these geometric structures in smaller blocks regarding the fibration, we were able to get an explicit expression of the dual $B$-field in terms of the starting data, as seen in (\ref{tilB}). Finally, the Theorem (\ref{theo}) provides the necessary constraints for the Topological dual of an emergent gravity to be an emergent gravity itself.

In section $4$ we showed that the constraints from Theorem (\ref{theo}) are not too restrictive by building examples of T-duality of emergent gravity: first on self T-dual tori, where the closeness of the $B$-fields are not an issue; and later on more complicated self T-dual nilmanifolds. In all these examples, the T-dual torus bundles have zero flux on both sides.

In section $5$ we discussed the difficulty of replicating such examples on T-dual torus bundles with non-zero flux. We showed that in the T-duality between the $4$-torus (with a non-zero flux) and the Kodaira-Thurston nilmanifold (with zero flux) there is no emergent gravity on one side that is T-dual to another emergent gravity on the other.

While all the computations and examples in this article are on nilmanifolds, we suspect that those can be extended to other classes of homogeneous spaces, such as flag manifolds, given the recent developments \cite{GGV}.

\end{document}